\documentclass[preprint,nofootinbib]{revtex4}

\usepackage{graphicx}
\usepackage{amsmath}
\usepackage{caption}
\usepackage{subcaption}
\usepackage{color}

\begin{document}

\title{Remnant loop quantum black holes}

\author{H. A. Borges$^1$, I. P. R. Baranov$^2$, F. C. Sobrinho$^{3}$ and S. Carneiro\footnote{Corresponding author}\footnote{saulo.carneiro.ufba@gmail.com}$^{1,4}$}

\affiliation{$^1$Instituto de F\'{\i}sica, Universidade Federal da Bahia, 40210-340, Salvador, BA, Brazil\\$^2$Instituto Federal de Educa\c c\~ao, Ci\^encia e Tecnologia da Bahia, 40301-015, Salvador, BA, Brazil\\$^3$Instituto de F\'isica, Universidade de S\~ao Paulo, 05508-090, S\~ao Paulo, SP, Brazil\\$^4$Observat\'orio Nacional, 20921-400, Rio de Janeiro, RJ, Brazil}

\date{\today}

\begin{abstract}

Polymer models inspired by Loop Quantum Gravity (LQG) have been used to describe non-singular quantum black holes with spherical symmetry, with the classical singularity replaced by a transition from a black hole to a white hole. A recent model, with a single polymerisation parameter, leads to a symmetric transition with same mass for the black and white phases, and to an asymptotically flat exterior metric. The radius of the transition surface is, however, not fixed, increasing with the mass. Following similar procedures, in a previous paper we have fixed that radius by identifying the minimal area on the transition surface with the area gap of LQG. This allowed to find a dependence of the polymerisation parameter on the black hole mass, with the former increasing as the latter decreases. It also permitted to extend the model to Planck scale black holes, with quantum fluctuations remaining small at the horizon. In the present paper we extend this analysis to charged black holes, showing that the Cauchy horizon lies beyond of the transition surface. We also show the existence of limiting states with zero surface gravity, the lightest one with $Q = 0$ and $m = \sqrt{2}/4$, and the heaviest with $Q = m = \sqrt{2}/2$. Using our solutions to approximate quasi-steady horizons, we show that Hawking evaporation leads asymptotically to these extremal states, leaving remnant black holes of Planck size. 

\end{abstract}

\maketitle

\section{Introduction}

The discussion about the ultimate fate of evaporating black holes is as old as difficult, as a full description of quantum black holes is still missing. We know that for a macroscopic black hole the temperature of the horizon scales with the inverse of its mass, which would eventually lead to its complete evaporation. However, the possibility of remnant Planck scale black holes has been considered on the basis of quantum gravity arguments \cite{Nature,Carr,chineses}. A promising formulation of quantum gravity is provided by Loop Quantum Gravity (LQG), a non-perturbative approach constructed on the space of holonomies \cite{lewandowski,livros1,livros2,livros3}. This theory motivated a treatment of the initial cosmological singularity in the realm of Loop Quantum Cosmology \cite{bojwald}, at the same time that provided an explanation for the Bekenstein-Hawking entropy of black hole horizons \cite{meissner,ghosh}. However, finding solutions in the scope of the full theory is a difficult task, and some effective models have been considered in the search for non-singular black holes. These make use of the polymerisation of classical Hamiltonians, which gives origin to a transition hypersurface in the horizon inner region, at which a tunneling from a black hole to a white hole takes place \cite{modesto,corichi,PRL,AOS,mariam,alemaes1,alemaes2,guillermo1,guillermo2,espanhois,bascos2,para}.

An interesting, uniparametric polymerisation scheme (ABBV from now on) has been recently used to obtain a symmetric tunneling, with black and white phases with equal masses \cite{espanhois,bascos2}. In contrast to previous procedures, the exterior metric is asymptotically flat, whereas the transition surface is inside the event horizon for any value of the polymerisation parameter. Nevertheless, as the latter is considered constant on the phase space, the value of the transition radius, $r_0$, is not fixed and increases with the black hole mass, loosing contact with its quantum gravity origin. A different treatment was proposed in \cite{PRL,AOS}, where the polymerisation parameters are promoted to conjugate variables, conserved along a given dynamical trajectory. In this way, they are functions of the black hole mass, decreasing with the mass, while the transition radius is fixed by imposing that the minimum area defined by holonomies on $2$-surfaces is proportional to the LQG minimal area\footnote{This procedure is not free of criticisms, see for example \cite{referee2,guillermo3}. For a discussion on the covariance of polymer models, see \cite{bojowald}.}.

A similar constraint was also imposed in the ABBV context \cite{Fernando}, which allowed the model extension to the Planck regime. In particular, the horizon area of microscopic black holes has the same mass dependence as in the classical case, which results from the fact that the radial coordinate is not polymerised and, in consequence, the angular part of the metric is not modified. Furthermore, it was shown that, at this scale, the curvature scalars are negligible at the horizon as compared to the transition surface, with relatively small corrections with respect to the classical values. This is valid for neutral, spherically symmetric black holes, but the functional coincidence between the horizon areas of the classical and quantum solutions seems to be valid also for rotating black holes, at least in the extremal case \cite{FoP}. In fact, a horizon with Planck mass and angular momentum $\hbar$ has a classical area corresponding exactly to a surface pierced by four spin network lines of colour $j=1/2$, provided that the Barbero-Immirzi parameter is $\gamma = \sqrt{3}/6$. This is $5\%$ above the approximate value derived from the Bekenstein-Hawking entropy in the limit of large horizons \cite{meissner,ghosh} and precisely gives, for Planck scale horizons, the correct leading order term in the entropy versus area relation \cite{CQG}.

The main goal of the present paper is to verify whether extremal horizons are possible in the spherical case. It is possible to generalise the ABBV polimerisation to charged black holes, and in Ref.~\cite{bascos3} an exhaustive analysis is made in a cosmological setup. Here we will apply a minimal area condition by imposing that the $2$-sphere of radius $r_0$ has the LQG minimum area \cite{modesto}. This is simpler than the pair of conditions imposed in \cite{PRL,AOS}, and such a simplification is possible because in the present model only one of the two canonical variables is polymerised. Under this assumption, we find a simple relation between the polymerisation parameter and the black hole mass, with the former increasing as the latter decreases. We then show that there is a finite, minimal mass for which the event horizon and the transition surface coincide, while the surface gravity vanishes. In the context of an evaporation process, it is an asymptotic state, reached in an infinite time. For an asymptotic observer it represents a remnant black hole, with Komar energy equal to that minimal mass. Surprisingly enough, such extremal states occur both for charged holes and in the Schwarzschild case.

\section{The model}

In the canonical variables used in \cite{PRL,AOS}, the classical Reissner-Nordstr\"om Hamiltonian \cite{bascos3,rakesh,esteban} can be written as
\begin{equation}\label{eq:hcl}
    H_{\rm cl}=-\frac{1}{2 G \gamma}\left[\left(b+\frac{\gamma^2}{b}-\frac{\gamma^2Q^2}{b p_c} \right) p_b + 2c p_c    \right],
\end{equation}
{where the conjugate variables $b$, $p_b$, $c$ and $p_c$ obey the algebra $\{b,p_b\} = G\gamma$ and $\{c,p_c\} = 2G\gamma$,
with associate homogeneous metric}
\begin{equation} \label{homogeneous}
    ds^2 = - N^2 dT^2 + \frac{p_b^2}{p_c} dx^2 + p_c d\Omega^2,
\end{equation}
with the lapse given by $N = \gamma \sqrt{p_c}/b$. {These canonical variables can be promoted to quantum operators through a polimerisation procedure, defined in our case by the transformations \cite{florencia}}
\begin{equation}\label{eq:polymerisation}
b \rightarrow \dfrac{\sin\left(\delta_b b\right)}{\delta_b}, \quad \quad \quad p_b \rightarrow \dfrac{p_b}{\cos(\delta_b b)}.
\end{equation}
{where $\delta_b$ is the polimerisation parameter. The classical limit is recovered by doing $\delta_b \rightarrow 0$. Then, by also introducing the regularisation factor}
\begin{equation}\label{eq:regularisation}
\dfrac{\cos(\delta_b b)}{\sqrt{1+\gamma^2\delta_b^2}},
\end{equation}
{we can construct the effective Hamiltonian}
\begin{align}\label{eq:abbvhamiltonian}
H_{\rm eff}
&= -\dfrac{1}{2G\gamma\sqrt{1+\gamma^2\delta_b^2}}\left[  \left(\dfrac{\sin(\delta_b b)}{\delta_b} +\frac{\gamma^2 \delta_b}{\sin{(\delta_b b)}}- \dfrac{\gamma^2 \delta_b Q^2}{\sin( \delta_b b)p_c }\right)p_b + 2cp_c\cos(\delta_b b)\right].
\end{align}
The dynamical equations are
\begin{align}
\dot{b} &= \{b, H_{\rm eff}\} = G\gamma\dfrac{\partial H_{\rm eff}}{\partial p_b} = -\dfrac{1}{2\sqrt{1+\gamma^2\delta_b^2}}\left(\dfrac{\sin(\delta_b b)}{\delta_b} + \dfrac{\delta_b\gamma^2}{\sin(\delta_b b)} - \frac{\gamma^2 \delta_b Q^2}{\sin (\delta_b b) p_c}\right)\label{eq:bponto},\\
\dot{c} &=\{c, H_{\rm eff}\}= 2G\gamma\dfrac{\partial H_{\rm eff}}{\partial p_c} = -\dfrac{2c\cos(\delta_b b)}{\sqrt{1+\gamma^2\delta_b^2}}-\dfrac{\gamma^2 \delta_b Q^2}{\sin( \delta_b b)}\frac{p_b}{p_c^2 }\label{eq:cponto},\\
\dot{p_b} &=\{p_b, H_{\rm eff}\}= -G\gamma\dfrac{\partial H_{\rm eff}}{\partial b}
\nonumber
\\
&= \dfrac{1}{2\sqrt{1+\gamma^2\delta_b^2}}\left[-2cp_c\sin(\delta_b b)\delta_b + \left(1 - \dfrac{\delta_b^2\gamma^2}{\sin^2(\delta_b b)}+\frac{\gamma^2 \delta_b^2 Q^2}{p_c \sin^2{(\delta_b b)}}\right)p_b\cos(\delta_b b)\right]\label{eq:pponto},\\
\dot{p_c} &=\{p_c, H_{\rm eff}\}= -2G\gamma\dfrac{\partial H_{\rm eff}}{\partial c} = \dfrac{2p_c\cos(\delta_b b)}{\sqrt{1+\gamma^2\delta_b^2}},\label{eq:pcponto}
\end{align}
where the dot means derivative with respect to the time variable $T$.

It can be shown that
\begin{equation}\label{eq:b}
    \frac{\sin^2(\delta_b b)}{\gamma^2 \delta_b^2}=\frac{2m}{\sqrt{p_c}}-1-\frac{Q^2}{p_c}
\end{equation}
is solution of Eq.~\eqref{eq:bponto}, where $m$ is a constant of motion that will be identified with the black hole mass. Substituting into \eqref{eq:pcponto} we obtain
\begin{equation}\label{eq:pc}
    p_c(T)=\frac{e^{-2 T}}{{4 b_0^4 (b_0+1)^2  m^2}}{\Bigl\{(b_0+1 )m^2 \bigl[ b_0-1 + ( b_0+1) e^T\bigr]^2  -  \
(b_0-1 ) b_0^2 Q^2\Bigr\}^2},
\end{equation}
where
\begin{equation} \label{b0}
b_0 \equiv \sqrt{1+\gamma^2 \delta_b^2}
\end{equation}
and the integration constant was chosen so that, for $Q=0$, we recover the uncharged solution \cite{Fernando}.
We can obtain the radius of the transition surface by doing $\dot{p}_c=0$. For $b_0 = 1$ (the classical case) we have $p_{c}^{\rm min} = 0$, while for $b_0 > 1$ the only real solution is given by
\begin{equation} \label{r0}
    r_0 = \sqrt{p_{c}^{\rm min}}=\frac{(b_0^2-1)}{b_0^2}m \left(1+\sqrt{1-\frac{b_0^2 Q^2}{(b_0^2-1)m^2}} \right),
\end{equation}
with
\begin{equation} \label{Qmax}
|Q|\leq m \frac{\sqrt{b_0^2-1}}{b_0}.
\end{equation}

Solving the remaining equations of motion and using a suitable change of variables, we obtain, by analytic continuation, the static metric
\begin{equation} \label{metric}
    ds^2 = -\left( 1 - \frac{2m}{r} + \frac{Q^2}{r^2} \right) d\tau^2 + \left( 1 - \frac{2m}{r} + \frac{Q^2}{r^2} \right)^{-1} \left( 1 - \frac{r_0}{m} g(r) \right)^{-1} dr^2 + r^2 d\Omega^2,
\end{equation}
where $r \equiv \sqrt{p_c}$ and
\begin{equation}
    g(r) = \frac{\frac{2m}{r} - \frac{Q^2}{r^2}}{1 + \sqrt{1 - \frac{b_0^2 Q^2}{(b_0^2-1)m^2}}}.
\end{equation}
This metric is asymptotically flat, and presents the same horizons as in the classical theory, with radii
\begin{equation} \label{horizons}
    r_h^{\pm} = m \left( 1 \pm \sqrt{1 - \frac{Q^2}{m^2}} \right).
\end{equation}
It is possible to show that the metric tensor component $g_{rr}$ also diverges for $r=r_0$.

\section{Extremal horizons}

For having $\delta_b$ as a function of $m$ we apply the minimal area condition \cite{modesto}
\begin{equation} \label{minimal_area}
    4\pi p_{c}^{\rm min} = 4 \pi \sqrt{3} \gamma,
\end{equation}
which from (\ref{r0}) leads to\footnote{From now on we will fix $\gamma = \sqrt{3}/6$, but our conclusions do not depend on this particular choice.}
\begin{equation} \label{delta_b}
    \delta_b^2 = \frac{12}{2\sqrt{2}m - 2Q^2 -1}.
\end{equation}
{Since $m$ and $Q$ are constants of motion, we see that $\delta_b$ is also a constant of motion, as assumed when we derived the Hamilton equations from (\ref{eq:abbvhamiltonian}). The constraint (\ref{delta_b}) selects, from all the possible dynamical trajectories, only those that match the LQG minimal area (\ref{minimal_area}) at the transition surface}.

Combining (\ref{delta_b}) with (\ref{Qmax}) we see that, for a given mass, the black hole charge is limited by
\begin{equation} \label{Qmax2}
  Q^2 \leq \frac{\sqrt{2}}{2} m.
\end{equation}
From (\ref{delta_b}) we also have
\begin{equation} \label{limite}
    2\sqrt{2}m > 2Q^2 + 1.
\end{equation}
Therefore, from (\ref{horizons}) we can show that
\begin{equation}
    r_h^- < r_0 < r_h^+,
\end{equation}
that is, the minimal area condition implies that the Cauchy horizon lies always beyond of the transition surface and is never reached. 
For charges saturating the limit (\ref{limite}), the polymerisation parameter $\delta_b$ diverges, $b_0 \gg 1$ and $r_h^+ \rightarrow r_0$, while
\begin{equation}
    r_h^- = 2m - \frac{\sqrt{2}}{2}.
\end{equation}
In particular, for $Q=0$ we have $r_h^- = 0$ and $r_h^+/2 = m = \sqrt{2}/4$, whereas for the maximal charge (\ref{Qmax2}) we have $r_h^{\pm} = Q = m = \sqrt{2}/2$. 

\begin{figure}
\centering
\includegraphics[width=0.7\linewidth]{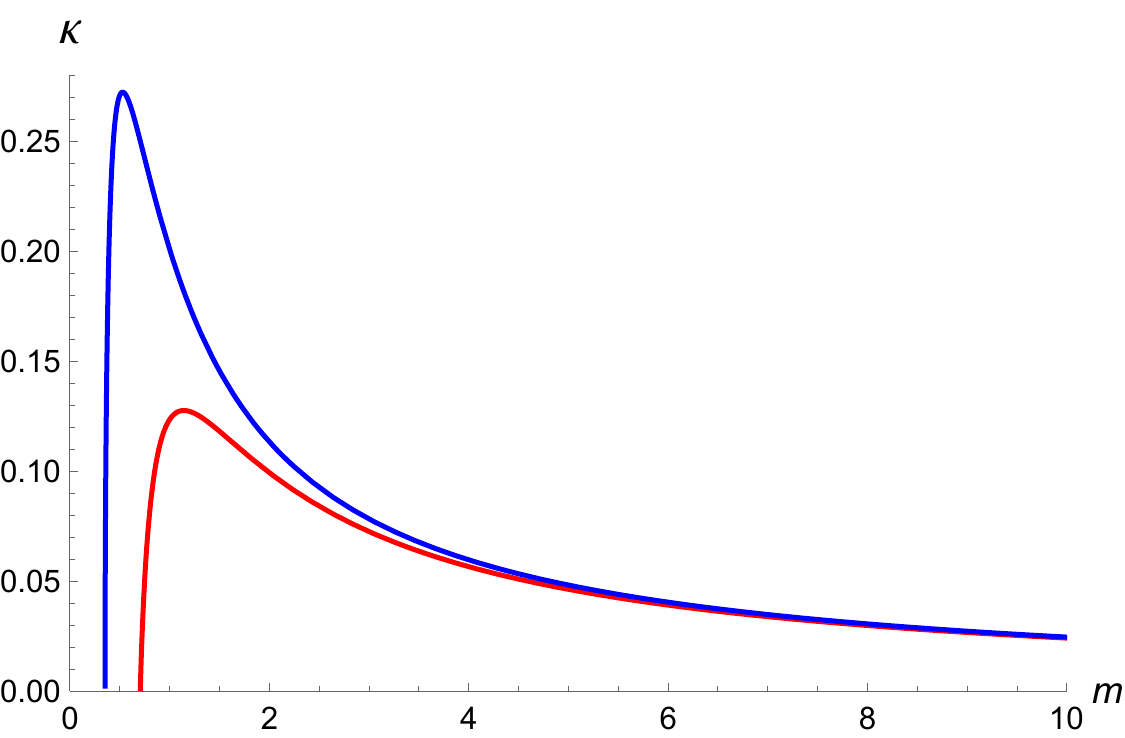}
\caption{The surface gravity as a function of the black hole mass for the maximally charged (red) and neutral (blue) solutions.}
\label{fig1}
\end{figure}

{The calculation of the Komar energy is straightforward \cite{wald}}, leading to
\begin{equation} \label{Komar}
    E(r) = \left( m - \frac{Q^2}{r} \right) \sqrt{1 - \frac{r_0}{m} g(r)}.
\end{equation}
For $r \rightarrow \infty$ we have $E = m$, while at the event horizon $E = r_h^2 \kappa$, {where $\kappa$ is the surface gravity \cite{wald}}. It is not difficult to show that $E(r_0)=0$. That is, when the inequality (\ref{limite}) is saturated, the Komar energy at the event horizon and the surface gravity vanish. In Fig.~\ref{fig1} we plot the surface gravity as a function of $m$ for the maximal charge (\ref{Qmax2}) (red) and for $Q = 0$ (blue).

We see from this figure that, while for macroscopic horizons the surface gravity increases when the black hole loses energy, when approaching the Planck scale this behaviour is reversed, with $\kappa$ going to zero as the mass is close to its minimum value. This has two important consequences. The first is that, near the minimal mass, we can treat Hawking evaporation as an adiabatic process, taking the present solutions as good approximations for quasi-static horizons. Second, it means that the evaporation leaves, asymptotically, a remnant Planck scale black hole with zero temperature and minimal mass. In this process, the horizon approaches the transition surface and the Komar energy goes to zero at the horizon, but remaining $E = m$ at spatial infinity.

We can show that the approach to the extremal state is indeed asymptotic by evaluating the evaporation time from an initial state with $E = E_i$, measured by a stationary observer. Dismissing gray factors, we have
\begin{equation}
    \frac{1}{r_h^2} \frac{dE}{d\tau} \propto (-\kappa^4) = - \left(\frac{E}{r_h^2}\right)^4.
\end{equation}
Close to the minimal mass, the horizon radius is nearly constant and we obtain
\begin{equation}
    \Delta \tau \propto \left( - r_h^{6} \right) \int_{E_{i}}^0 \frac{dE}{E^4} \rightarrow \infty.
\end{equation}

\begin{figure}
     \centering
     \begin{subfigure}[b]{0.45\textwidth}
         \centering
         \includegraphics[width=\textwidth]{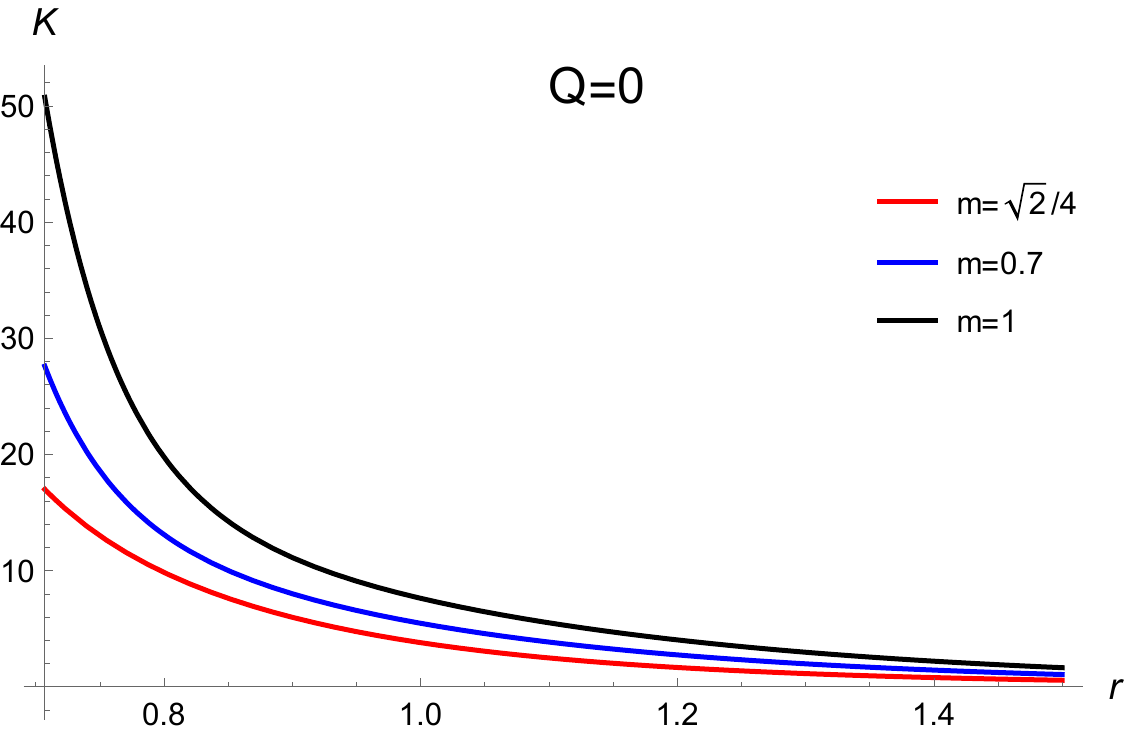}
         \label{fig:Qzero}
     \end{subfigure}
     \hfill
     \begin{subfigure}[b]{0.45\textwidth}
         \centering
         \includegraphics[width=\textwidth]{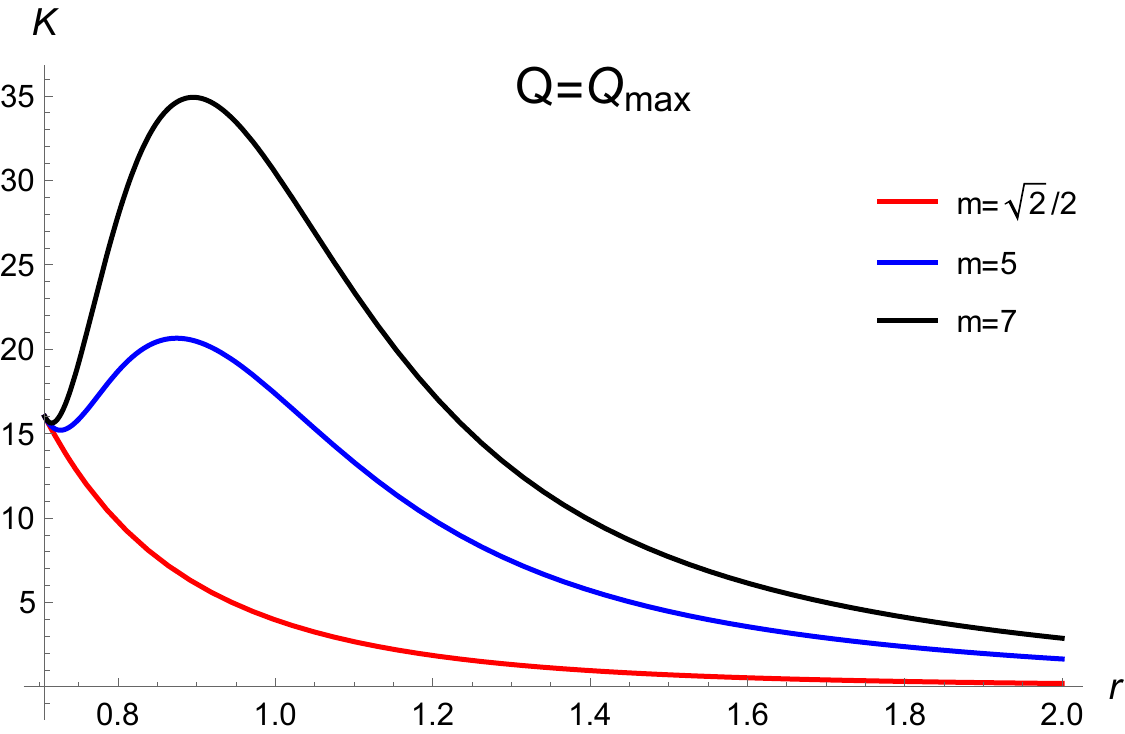}
         \label{fig:Qmax}
     \end{subfigure}
     \hfill
        \caption{Kretschmann scalar as a function of the radial coordinate and the black hole mass, for $Q=0$ (left) and for the maximal allowed charge (right).}
        \label{fig:Kretschmann}
\end{figure}

Finally, we have computed the Kretschmann scalar -- {defined from the Riemann tensor as $K=R^{\alpha \beta \gamma \delta}R_{\alpha \beta \gamma \delta}$} -- for our solutions. As shown in Fig.~\ref{fig:Kretschmann}, it is real and regular for $r \geq r_0$, with a maximum at $r_0$ for $Q=0$ (left panel). Such a maximum increases with the mass, with a minimum value at the extremal state $m = \sqrt{2}/4$, when the horizon coincides with the transition surface. For the maximal charge (\ref{Qmax2}) (right panel), the Kretschmann scalar at $r_0$ does not depend on the black hole mass. It remains finite in the extremal case $m = \sqrt{2}/2$, with a maximum at the transition surface. For large masses the maximum occurs at $r \approx 0.91$.

\section{The entropy balance}

The evaporation of classical black holes is the origin of a difficulty commonly known as ``the information paradox", related to the balance of entropy between the evaporating horizon and the emitted radiation. The horizon entropy is a quarter of its area and, when the black hole is completely evaporated, its entropy variation is $S = {\cal A}/4$. On the other hand, the produced thermal radiation carries an indefinitely large entropy, which would imply a lost of information in the whole system. Remnant black holes are a potential way to understand this apparent deficit of information, if all the initial degrees of freedom remain stored inside the relic horizon.

In the present model, this possibility can be investigated by writing the variation in the Komar energy inside a neutral, spherically symmetric horizon as
\begin{equation}
    dE = T dS = \frac{\kappa}{2\pi} dS = \frac{E}{2\pi r_h^2} dS,
\end{equation}
with $r_h = 2m$. From (\ref{Komar}) we have
\begin{equation}
    E (r_h) = m \sqrt{1 - \frac{r_0}{2m}},
\end{equation}
and, by using  ${\cal A} = 16 \pi m^2$, we obtain
\begin{equation}
    dS = \frac{d{\cal A}}{4} \left[ 1 + \frac{r_0}{4m \left(1 - \frac{r_0}{2m} \right) } \right].
\end{equation}
Its integration gives
\begin{equation}
    S = \frac{{\cal A}}{4} + \pi r_0 \left[ (r_h - r_0) + r_0 \ln (r_h - r_0) \right].
\end{equation}

For $r_h \gg r_0$ we have, apart an additive constant and fixing $r_0 = \sqrt{2}/2$,
\begin{equation}
    S = \frac{{\cal A}}{4} + \frac{\sqrt{2\pi}}{4} {\cal A}^{1/2} + \frac{\pi}{4} \ln {\cal A} \quad \quad (r_h \gg r_0),
\end{equation}
which reduces to the Bekenstein-Hawking entropy in the limit of large horizons. In the opposite limit $r_h \approx r_0$, the leading term is
\begin{equation}
    S = \frac{\pi}{2} \ln (r_h - r_0) \quad \quad (r_h \approx r_0).
\end{equation}
When $r_h \rightarrow r_0$, $S \rightarrow -\infty$. As the final stages of evaporation are adiabatic, with $T \approx 0$, we in fact expect that the black hole entropy variation balances that carried by radiation.

\section{Concluding remarks}

The modeling of dark matter as relic Planckian black holes is as old as the discussion about their stability \cite{Nature,modesto2,PLB,Nelson,referee1,theodor}. The formation of Planck mass black holes at the end of the inflationary phase has been recently investigated as resulting from the collisions of high energy particles \cite{PLB,Nelson}. For a reheating temperature of order $10^{17}$ GeV,\footnote{{This is above the limit of $10^{16}$ GeV imposed by CMB observations in the simpler cases of single-field slow-roll inflation. However, this limit can be respected if the particles chemical potential at this temperature is high enough \cite{PLB}.}} the number of Planck energy particles in the tail of a Fermi distribution is enough to have today the amount of dark matter observed in the galactic halos. {The inelastic scattering of repulsive charges at Planck energies can give origin to black holes of Planck mass if the total charge involved is $Q \sim 1$ \cite{Nelson}.} The same may happen in the scattering of Dirac neutrinos, because of the repulsion between their magnetic moments \cite{PLB}. {Ultralight primordial black holes may also be associated to other interesting phenomena, for example an alleviation of the current tension in the Hubble constant measurement \cite{1,2} or a possible solution for the baryon asymmetry problem \cite{3,4,5}, in both cases due to their Hawking evaporation. The existence of an early matter-dominated era triggered by microscopic black holes has been also investigated \cite{6}, giving rise to a rich gravitational wave phenomenology \cite{7,8}.}

{On the other hand, non-singular spherically symmetric solutions with black-to-white transition have recently been modeled in quantum gravity \cite{modesto,corichi,PRL,AOS,mariam,alemaes1,alemaes2,guillermo1,guillermo2,espanhois,bascos2,Rovelli1,Rovelli2}, including charged solutions \cite{bascos3,Rovelli3}. Compared to previous approaches, we can point out some interesting features of the polimerisation scheme adopted in \cite{espanhois,bascos2}. Among them, the symmetry between the black and white phases, which present the same mass; the asymptotic freedom presented by the external metric; and finally the formal coincidence of the horizon area with the corresponding classical solution.}

In the present letter we have modeled charged quantum black holes with the help of this suitable polymerisation, and by postulating the LQG area gap as the minimum allowed area for the transition surface. Our main result is the prediction, in the realm of such a model, of extremal states with Planck scale masses and zero surface gravity. 
Considering our solutions as approximations to quasi-steady states, we have computed the time of evaporation, showing that these remnants are reached asymptotically. The maximally charged state has $Q = \sqrt{2}/2 \approx 0.7$, four times the low-energy charge of a couple of electrons, $2e \approx 0.17$. Taking into account, on one hand, that we have not included the effect of vacuum polarisation in the charge estimation referred above and, on the other hand, the effective character of the present model, this approximate coincidence may not be fortuitous. In any case, it suggests that the search for extremal solutions in full or effective quantum gravity theories is an interesting route for understanding both the fate of evaporating black holes and the origin of dark matter.

\section*{Acknowledgements}

We are thankful to A. Saa and H. P. Costa for helpful discussions. FCS is supported by CAPES (Brazil). SC is partially supported by CNPq (Brazil) with grant 311584/2020-9.


\begin{thebibliography}{}

\bibitem{Nature} J. H. MacGibbon, Nature {\bf 329} (1987) 308.

\bibitem{Carr} X. Calmet, B. Carr and E. Winstanley, {\it Quantum Black
Holes} (Springer, 2014).

\bibitem{chineses} C. Zhang, Y. Ma, S. Song and X. Zhang, Phys. Rev. {\bf D102} (2020) 041502(R).

\bibitem{lewandowski} A. Ashtekar and J. Lewandowski, Class. Quantum Grav. {\bf 21} (2004) R53.

\bibitem{livros1} T. Thiemann, {\it Modern Canonical Quantum General Relativity} (Cambridge University Press, 2008).

\bibitem{livros2} R. Gambini and J. Pullin, {\it A first course in Loop Quantum Gravity} (Oxford University Press, 2011).

\bibitem{livros3} C. Rovelli and F. Vidotto, {\it Covariant Loop Quantum Gravity} (Cambridge University Press, 2015).

\bibitem{bojwald} M. Bojowald, Living Rev. Relativ. {\bf 11} (2008) 4.

\bibitem{meissner} K. A. Meissner, Class. Quantum Grav. {\bf 21} (2004) 5245.

\bibitem{ghosh} A. Ghosh and P. Mitra, Phys. Lett. {\bf B616} (2005) 114.

\bibitem{modesto} L. Modesto, Int. J. Theor. Phys. {\bf 49} (2010) 1649.

\bibitem{corichi} A. Corichi and P. Singh, Class. Quantum Grav. {\bf 33} (2016) 055006.

\bibitem{PRL} A. Ashtekar, J. Olmedo and P. Singh, Phys. Rev. Lett. {\bf 121} (2018) 241301.

\bibitem{AOS} A. Ashtekar, J. Olmedo and P. Singh, Phys. Rev.
{\bf D98} (2018) 126003.


\bibitem{mariam} M. Bouhmadi-L\'opez {\it et al.}, Phys. Dark Univ. {\bf 30} (2020) 100701.

\bibitem{alemaes1} N. Bodendorfer, F. M. Mele and J. M\"unch, Class. Quantum Grav. {\bf 36} (2019) 195015.

\bibitem{alemaes2} N. Bodendorfer, F. M. Mele and J. M\"unch, Class. Quantum Grav. {\bf 38} (2021) 095002.

\bibitem{guillermo1} B. Elizaga Navascués, A. García-Quismondo and G. A. Mena Marugán, Phys. Rev. {\bf D106} (2022) 063516.

\bibitem{guillermo2} B. Elizaga Navascués, A. García-Quismondo and G. A. Mena Marugán, Phys. Rev. {\bf D106} (2022) 043531.

\bibitem{espanhois} A. Alonso-Bardaji, D. Brizuela and R. Vera, Phys. Lett. {\bf B829} (2022) 137075.

\bibitem{bascos2} A. Alonso-Bardaji, D. Brizuela and R. Vera, Phys. Rev. {\bf D106} (2022) 024035.

\bibitem{para} Z. S. Moreira, H. C. D. Lima Junior, L. C. B. Crispino and C. A. R. Herdeiro, Phys. Rev. {\bf D107} (2023) 104016.

\bibitem{referee2} N. Bodendorfer, F. M. Mele and and J. M\"unch, Class. Quantum Grav. {\bf 36} (2019) 187001.

\bibitem{guillermo3} A. Garcia-Quismondo and G. A. Mena Marug\'an, Phys. Rev. {\bf D106} (2022) 023532.

\bibitem{bojowald} M. Bojowald, Phys. Rev. {\bf D103} (2021) 126025.

\bibitem{Fernando} F. C. Sobrinho, H. A. Borges, I. P. R. Baranov and S. Carneiro, Class. Quantum Grav. {\bf 40} (2023) 145003.

\bibitem{FoP} S. Carneiro, Found. Phys. {\bf 50} (2020) 1376.

\bibitem{CQG} C. Pigozzo, F. S. Bacelar and S. Carneiro, Class. Quantum Grav. {\bf 38} (2021) 045001.

\bibitem{bascos3} A. Alonso-Bardaji, D. Brizuela and R. Vera, Phys. Rev. {\bf D107} (2023) 064067.

\bibitem{rakesh} R. Tibrewala, Class. Quantum Grav. {\bf 29} (2012) 235012.

\bibitem{esteban} R. Gambini, E. M. Capurro and J. Pullin, Phys. Rev. {\bf D91} (2015) 084006.

\bibitem{florencia} R. Gambini, F. Ben\'itez and J. Pullin, Universe {\bf 8} (2022) 526.



\bibitem{wald} R. Wald, {\it General Relativity} (The University of Chicago Press, 1984).

\bibitem{modesto2} L. Modesto and I. Pr\'emont-Schwarz, Phys. Rev. {\bf D80} (2009) 064041.

\bibitem{PLB} S. Carneiro, P. C. de Holanda and A. Saa, Phys. Lett. {\bf B822} (2021) 136670.

\bibitem{Nelson} I. J. Araya {\it et al.}, JCAP {\bf 2302} (2023) 030.

\bibitem{referee1} S. Kazemian, M. Pascual, C. Rovelli and F. Vidotto, Class. Quantum Grav. {\bf 40} (2023) 087001.

\bibitem{theodor} T. Papanikolaou, Class. Quantum Grav. {\bf 40} (2023) 134001.

\bibitem{1} D. Hooper, G. Krnjaic and S. D. McDermott, JHEP {\bf 08} (2019) 001.

\bibitem{2} T. Papanikolaou, arXiv:2303.00600 (2023).

\bibitem{3} J. D. Barrow, E. J. Copeland, E. W. Kolb and A. R. Liddle,
Phys. Rev. {\bf D43} (1991) 984.

\bibitem{4} N. Bhaumik, A. Ghoshal and M. Lewicki, JHEP {\bf 07} (2022) 130.

\bibitem{5} T. C. Gehrman, B. S. E. Haghi, K. Sinha and T. Xu, JCAP {\bf 02} (2023) 062.

\bibitem{6} J. Garcia-Bellido, A. D. Linde and D. Wands, Phys. Rev. {\bf D54} (1996) 6040.

\bibitem{7} T. Papanikolaou, V. Vennin and D. Langlois, JCAP 03 (2021) 053.

\bibitem{8} G. Domenech, C. Lin and M. Sasaki, JCAP {\bf 04} (2021) 062.

\bibitem{Rovelli1} H. M. Haggard and C. Rovelli, Phys. Rev. {\bf D92} (2015) 104020.

\bibitem{Rovelli2} E. Bianchi {\it et al.}, 
Class. Quantum Grav. {\bf 35} (2018) 225003.

\bibitem{Rovelli3} A. Rignon-Bret and C. Rovelli, Phys. Rev. {\bf D105} (2022) 086003.


\end{thebibliography}
\end{document}